\documentclass[twocolumn,showpacs,preprintnumbers,showkeys,superscriptaddress]{revtex4-1}
\usepackage{CJK}
\usepackage{mathrsfs}
\usepackage{amssymb}
\usepackage{amsmath}
\usepackage{graphicx}
\usepackage{dcolumn}
\usepackage{bm}
\usepackage{graphicx}
\usepackage{float}
\usepackage[normalem]{ulem}
\usepackage{color}
\usepackage{multirow}
\usepackage{balance}
\begin{document}

\begin{CJK*}{UTF8}{} 

\balance

\title{$(h_{11/2})^2$ alignments in neutron-rich $^{132}$Ba with negative-parity pairs}

\author{Y. Lei ({\CJKfamily{gbsn}雷杨})}
\email{leiyang19850228@gmail.com}
\affiliation{Key Laboratory of Neutron Physics, Institute of Nuclear Physics and Chemistry, China Academy of Engineering Physics, Mianyang 621900, China}

\author{Z. Y. Xu ({\CJKfamily{gbsn}徐正宇})}
\affiliation{INPAC, Physics Department, Shianghai Jiao Tong University, Shianghai 200240, China}

\date{\today}

\begin{abstract}
The shell-model collective-pair truncation with negative-parity pairs is adopted to study the $(h_{11/2})^2$ alignment in $^{132}$Ba. The proton $(h_{11/2})^2$-alignment state is predicted as an $E\sim4.6$ MeV and $\tau\sim 0.5~\mu$s isomer with relatively strong E3 decay channels. The oblately deformed neutron $(h_{11/2})^{-2}$ alignment in the yrast band and four negative-parity bands are confirmed, even although two of these negative-parity bands favor the prolate deformation, which directly manifests the $\gamma$ unstability of $^{132}$Ba.
\end{abstract}
\maketitle
\end{CJK*}
$^{132}$Ba is a typical $\gamma$-unstable nucleus in the transitional region of $Z\geqslant50$ and $N\leqslant82$ with a large variety of coexisting nuclear shapes. Its negative-parity bands with $I^{\pi}=5^-$ and $6^-$ bandheads (bandhead energies: 2.12 and 2.358 MeV, respectively) maintain an oblate shape, while prolately deformed negative-parity bands with $I^{\pi}=7^-$ and $8^-$ bandheads (bandhead energies: 2.902 and 3.105 MeV, respectively) were also reported in the $^{132}$Ba level scheme \cite{neg-band-1,neg-band-2}. There four negative-parity bands here are denoted by ``$5^-$, $6^-$, $7^-$ and $8^-$ bands" in sequence. In the yrast band, the $\tau=$8.94(14) ns 10$^+$ isomer is assigned as the neutron $(h_{11/2})^{-2}$ alignment with an oblate shape, which has two $h_{11/2}$ neutron holes rotate alignedly. On the other hand, the proton $(h_{11/2})^2$ alignment with the prolate deformation is also expected in $^{132}$Ba (unfortunately unobserved yet), because the competition between proton and neutron $(h_{11/2})^{2}$ alignments are generally exhibited in this nuclear region \cite{ps-1,ps-2,ps-3,ps-4,ps-5,ps-6,ps-7,ps-8,ps-9,ps-10}. 

There may exist strong electromagnetic transitions from the $(\pi h_{11/2})^2$-alignment state to states in $7^-$ and $8^-$ bands, because initial and final states share a similar prolate shape. It's desirable to theoretically estimate these transition rates before an experimental search for the $(\pi h_{11/2})^2$ alignment in $^{132}$Ba. 

The $(\nu h_{11/2})^{-2}$ alignment was also suggested in $5^-$, $6^-$, $7^-$ and $8^-$ bands according to their band irregularity around 5$\sim$6 MeV \cite{neg-band-1,neg-band-2}. However, the experimental evidence is not as solid as the $(\nu h_{11/2})^{-2}$ alignment in the yrast $10^+$ isomer \cite{gfactor1,gfactor2}. It's essential to confirm the $(\nu h_{11/2})^{-2}$ alignment in these negative-parity bands from a shell-model perspective.

The present work aims at studying the $(h_{11/2})^2$ alignment in both positive and negative-parity states of $^{132}$Ba within the shell-model framework. The model-space truncation is required to reduce the gigantic dimension of the shell-model description for $^{132}$Ba. It's noteworthy that the collective-pair truncation of the Shell Model has be proved to be an efficient approach for the $(h_{11/2})^2$-alignment description \cite{yoshinaga-back-ce, yoshinaga-back-xe, yoshinaga-back-ba}. In such truncation, collective pairs with spin 0$\hbar$ and 2$\hbar$ are normally employed to represent the low-lying collectivity \cite{npa-cal-sd1,npa-cal-sd2,npa-cal-sd3,npa-cal-sd4,npa-cal-sd7,npa-cal-jiang-1,npa-cal-jiang-2}, which naturally provide $\Delta I=2$ band structures, as $5^-$, $6^-$, $7^-$ and $8^-$ bands of $^{132}$Ba behave. Recently, negative-parity pairs were introduced into the pair truncation \cite{n-74,e1-1,e1-2}, which enables a shell-model description of negative-parity states in a heavy even-nucleon system, e.g., negative-parity states of $^{132}$Ba discussed here. Thus, the collective-pair truncation with negative-parity pairs is adopted for our negative-parity-state related study on the $(h_{11/2})^2$ alignment of $^{132}$Ba.


\begin{table}
\caption{Hamiltonian parameters from Ref. \cite{yoshinaga-back-ba} in MeV.}\label{par}
\begin{tabular}{cccccccccccccccccccccccccccccccc}
\hline\hline
	&	   $s_{1/2}$   	&	   $d_{3/2}$   	&	   $d_{5/2}$   	&	   $g_{7/2}$   	&	   $h_{11/2}$	&		\\
\hline													
$\varepsilon_{\pi}$ 	&	   2.990   	&	   2.708   	&	   0.962   	&	   0.000   	&	   2.793   	&		\\
$\varepsilon_{\nu}$ 	&	   0.332   	&	   0.000   	&	   1.655   	&	   2.434   	&	   0.242   	&		\\
\hline													
$G_{\pi}^0$	&	$G_{\pi}^2$	&	$G_{\nu}^0$	&	$G_{\nu}^2$	&	$\kappa_{\pi}$	&	$\kappa_{\nu}$	&	$\kappa_{\pi\nu}$	\\
\hline													
0.130	&	0.030	&	0.130	&	0.026	&	0.045	&	0.065	&	0.070	\\
\hline\hline
\end{tabular}
\end{table}

Our calculation adopts a phenomenological shell-model Hamiltonian as in Ref. \cite{yoshinaga-back-ba}:
\begin{eqnarray}\label{ham}
H&=&-\sum\limits_{\sigma=\pi,~\nu}
(\sum\limits_j\varepsilon_{j\sigma}\hat{n}_{j\sigma} +
\sum\limits_{s=0,2}G^s_{\sigma}{\cal P}^{s\dag}_{\sigma}\cdot \tilde{\cal
P}^{s}_{\sigma} \nonumber\\
 &&+ \kappa_{\sigma} Q_{\sigma}\cdot Q_{\sigma} )  +\kappa_{\pi\nu}Q_{\pi}\cdot Q_{\nu},
\end{eqnarray}
with
\begin{eqnarray}\label{operator}
 &{\cal P}^{0\dagger}=   \sum\limits_{a} \frac{\sqrt{2{a}+1}}{2}(C_{a}^{\dagger} \times
C_{a}^{\dagger})^{0},\\
&{\cal P}^{2\dagger} = \sum\limits_{ab} q(a b) ( C^{\dagger}_{a} \times
C^{\dagger}_{b} )^{2},
Q = \sum\limits_{ab} q(a b) ( C^{\dagger}_{a} \times \tilde{C}_{b}
)^{2}. \nonumber
\end{eqnarray}
In Eq. (\ref{operator}), $q(ab) =\langle a||r^2Y^2|| b\rangle/r^2_0$, where $r_0$ is the oscillator parameter, $\sqrt{\hbar/(m\omega)}$. Hamiltonian parameters, i.e., $\varepsilon_{j\sigma}$, $G^S_{\sigma}$, $\kappa_{\sigma}$, and $\kappa_{\pi\nu}$ in Eq. (\ref{ham}), are also taken from Ref. \cite{yoshinaga-back-ba} as listed in Table \ref{par}.

Our pair-truncated shell-model space for $^{132}$Ba is given by the coupling of three proton pairs and three (hole-like) neutron pairs in the 50-82 shell. These pairs can be formally defined by
\begin{eqnarray}\label{pair}
A^{{I^{\pi}}\dagger} = \sum_{a\leqslant b} \beta^{I^{\pi}}_{ab} A^{{I^{\pi}}\dag} (ab),~
A^{{I^{\pi}}\dag}(ab) = \frac{( C^{\dagger}_a \times
C^{\dagger}_b )^{I^{\pi}}}{\sqrt{1+\delta_{ab}}},
\end{eqnarray}
where $C^{\dagger}_{a}$ is the creation operator of the $a$ orbit. Thus, $A^{{I^{\pi}}\dagger}(ab)$ is a non-collective pair with two nucleons at $a$ and $b$ orbits; $A^{{I^{\pi}}\dagger}$ represents a collective pair with optimized structure coefficients, $\beta^{I^{\pi}}_{ab}$, to best describe the nuclear low-lying collectivity. Especially, $\beta^{I^{\pi}=0^+}_{ab}$ corresponds to the Cooper-pair structure due to the strong nuclear pairing collectivity, and is determined following the principle of projected-particle-number BCS theory \cite{pbcs}. Other $\beta^{I^{\pi}}_{ab}$ with ${I^{\pi}}\neq0^+$ is obtained by the diagonalization in the $\nu=2$ broken-pair model space \cite{bpa}. More details about the determination of $\beta^{I^{\pi}}_{ab}$ are described in Ref. \cite{npa-cal-other3}.

In our calculation, four types of pairs are adopted:
\begin{itemize}
\item
[1] Following previous pair-truncation calculations \cite{npa-cal-sd1,npa-cal-sd2,npa-cal-sd3,npa-cal-sd4,npa-cal-sd7,npa-cal-jiang-1,npa-cal-jiang-2}, collective pairs with ${I^{\pi}}=0^+$ and $2^+$ are introduced. 
\item
[2] Bandbeads of $5^-$ and $6^-$ bands correspond to the mixture of neutron $h_{11/2}\times s_{1/2}$ and $h_{11/2}\times d_{3/2}$ configurations \cite{neg-band-1,neg-band-2}. In the 50-82 major shell, such mixture can only emerge in collective $I^{\pi}=5^-$ and $6^-$ neutron pairs. Thus, these two pairs are introduced to develop $5^-$ and $6^-$ bands, and denoted by ``$5^-$ and $6^-$" pairs.
\item
[3] $7^-$ and $8^-$ bands are built on the coupling of $h_{11/2}$ and $g_{7/2}$ protons \cite{neg-band-1,neg-band-2}. Considering the short-range property of nuclear force, the coupling with total spin $I=9$ should give the lowest binding energy \cite{casten}. Thus, the non-collective $(\pi h_{11/2} \times \pi g_{7/2})^{I^{\pi}=9^-}$ pair is introduced. 
\item
[4] We take the $(\pi h_{11/2} \times \pi h_{11/2})^{I^{\pi}=10^+}$ pair to describe the $(\pi h_{11/2})^2$ alignment.
\end{itemize}

\begin{table}
\caption{Structure coefficients, i.e., $\beta^{I^{\pi}}_{ab}$ defined in Eq. (\ref{pair}), of $5^-$ and $6^-$ pairs.}\label{pair-structure}
\begin{tabular}{lccccccccccccccccccccccccccccccc}
					\hline\hline
					&	$I^{\pi}=5^-$	&	$I^{\pi}=6^-$	\\
					\hline
$ab=h_{11/2}\times s_{1/2}$	&	$+$0.551	&	$+$0.616	\\
$ab=h_{11/2}\times d_{3/2}$	&	$-$0.832	&	$+$0.780	\\
$ab=h_{11/2}\times d_{5/2}$	&	$+$0.038	&	$+$0.097	\\
$ab=h_{11/2}\times g_{7/2}$	&	$-$0.050	&	$+$0.053	\\
					\hline\hline
				\end{tabular}
\end{table}

In previous pair-truncation calculations \cite{yoshinaga-back-xe,yoshinaga-back-ba,yoshinaga-back-ce}, the $(\nu h_{11/2})^{-2}$ alignment is represented by the $(\nu h_{11/2} \times \nu h_{11/2})^{I^{\pi}=10^+}$ pair. However, we don't introduce such pair, because the coupling of two $5^-$ and/or $6^-$ pairs with total angular momentum $I \geqslant 10\hbar$ (denoted by the ``$5^-\otimes 6^-$ coupling") can also describe the $(\nu h_{11/2})^{-2}$ alignment. To demonstrate this point, we list structure coefficients of $5^-$ and $6^-$ pairs in Table \ref{pair-structure}, which suggests that the $5^-\otimes 6^-$ coupling are mainly constructed by two $s_{1/2}$ and/or  $d_{3/2}$ neutrons along with two $h_{11/2}$ neutrons. These two $s_{1/2}$ and/or $d_{3/2}$ neutrons only provide little angular-momentum contribution, so that the $I \geqslant 10\hbar$ total angular momentum of the $5^-\otimes 6^-$ coupling almost comes from the rest two $h_{11/2}$ neutrons. In other words, these two $h_{11/2}$ neutrons have to contribute $I\sim 10\hbar$ angular momentum, which forces them to rotate alignedly. Thus, the  $(\nu h_{11/2})^{-2}$ alignment emerges within the $5^-\otimes 6^-$ coupling.

For $^{132}$Ba with 3 valence proton pairs and 3 valence neutron-hole pairs in 50-82 major shell, our proton or neutron pair-truncated basis is given by
\begin{eqnarray}
 |\tau_{\sigma=\pi,~\nu}^{J}\rangle=\left( \left( A^{r_1\dag} \times A^{r_2\dag} \right)^{(J_2)}
 \times A^{r_{3}\dag} \right)^{(J)}|0 \rangle,
 \end{eqnarray}
where $A^{r_1\dag}$, $A^{r_2\dag}$ and $A^{r_3\dag}$ can be any type of pairs we described above. Proton and neutron basis are coupled together as $|\tau^{J=J_{\pi}\times J_{\nu}}\rangle=|\tau_{\pi}^{J_{\pi}}\rangle\times |\tau_{\nu}^{J_{\nu}}\rangle$. Hamiltonian matrix elements under $|\tau^J\rangle$ bases can be calculated with the formalism of Ref. \cite{zhao-for}. These elements are denoted by $H^{\tau}_{ij}$, where $i$ and $j$ are indexes to identify bases in the diagonalization algorithm. The Hamiltonian should be diagonalized under a set of linearly independent, normalized and orthogonal bases, which can be constructed by the linear combination of $|\tau^J\rangle$ bases as described in Ref. \cite{chen-for}. In detail, we calculate the overlap matrix of $|\tau^J\rangle$ bases, and diagonalize it with a serial of eigenvalues ($e_i$) and corresponding eigenvector matrix elements ($v_{ij}$). The $i$th normalized and orthogonal basis is given by
\begin{equation}\label{linear_tran}
|\phi_i^J\rangle=\frac{1}{\sqrt{e_i}}\sum\limits_j v_{ij} |\tau_j^J\rangle.
\end{equation}
To handle the overcompleteness, we neglect $|\phi_i^J\rangle$s with $e_i=0$, and reserved $|\phi^J\rangle$ bases construct the pair-truncated space for $^{132}$Ba. Thus, the Hamiltonian matrix under $|\phi^J\rangle$ bases is produced by
\begin{equation}
H^{\phi}_{ij}=\frac{1}{\sqrt{e_ie_j}}\sum\limits_{kl}v_{ik}v_{jl}H^{\tau}_{kl},
\end{equation}
and inputted into Hamiltonian diagonalization. Resultant wave-functions are used for further calculations on electromagnetic properties and expectation values of the pair numbers.

The electromagnetic-transition operators adopted in our calculation are
\begin{eqnarray}\label{te}
T(E2)&=&\sum\limits_{\sigma=\pi,~\nu}e_{\sigma}r^2_{\sigma}Y^2_{\sigma},~T(E3)=\sum\limits_{\sigma=\pi,~\nu}e_{\sigma}r^3_{\sigma}Y^3_{\sigma},\\ \nonumber
T(M1)&=&\sqrt{\frac{3}{4\pi}}\sum\limits_{\sigma=\pi,~\nu}g_{l\sigma}\overrightarrow{l}_{\sigma}+g_{s\sigma}\overrightarrow{s}_{\sigma}.
\end{eqnarray}
Here, $e_{\sigma}$, $g_{l\sigma}$ and $g_{s\sigma}$ (in units of $e$ and $\mu_N/\hbar$) are the effective charge, orbital and spin gyromagnetic ratios of valence nucleons, respectively. They are taken from Ref. \cite{yoshinaga-back-ba} as $e_{\pi} = 2$, $e_{\nu} =-1$, $g_{\pi s} = 5.58\times0.7$, $g_{\nu s} =-3.82\times 0.7$, $g_{\pi l} = 1.05$ and $g_{\nu l }= 0.05$.


\begin{figure}
\includegraphics[angle=0,width=0.45\textwidth]{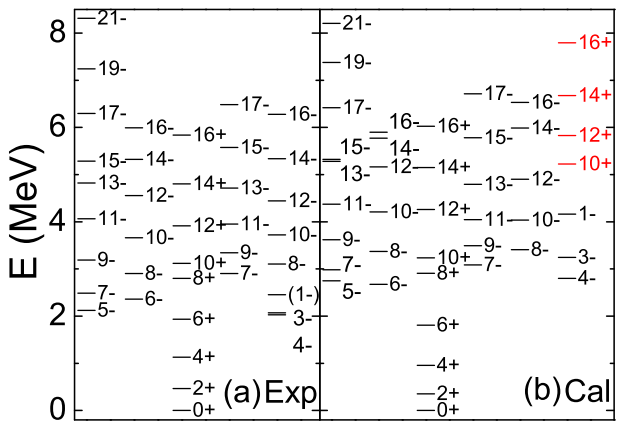}
\caption{(Color online) Experimental (a) \cite{neg-band-2,ensdf} and the calculated spectrum (b), including the yrast band, $5^-$, $6^-$, $7^-$, $8^-$ bands, and the $(\pi h_{11/2})^2$-alignment band, which is highlighted by red color.}\label{spe}
\end{figure}

Fig. \ref{spe} presents the calculated $^{132}$Ba spectrum compared with the experimental one, including the yrast band, $5^-$, $6^-$, $7^-$, $8^-$ bands and the potential $(\pi h_{11/2})^2$-alignment band. A rough spectral consistency between our calculation and the experiment is achieved for these bands we are concerned about. One may also note that there is still inconsistency for low-lying $1^-$, $3^-$ and $4^-$ states, which are all out of our scientific motivation and our model space constructed with four types of pairs we itemize above \cite{neg-band-2,oct,sm-failure}. Thus, we have no intention to fix the inconsistency by introducing redundant pairs, which obviously will complicate our study, and in what follows we limit ourselves to current model space. Furthermore, calculated $7^-$ and $8^-$ bands based on the $(\pi h_{11/2}\times \pi g_{9/2})^{I^{\pi}=9^-}$ pair are still observed to be systematically higher than those from experiments by $\sim 0.3$ MeV. This is because the  single-particle energy of the $^{132}$Ba $h_{11/2}$ proton, i.e., the $[550]\frac{1}{2}^-$ level, is depressed by $\sim 0.3$MeV relatively to the Fermi surface \cite{nillson}, due to the $\beta_2\sim 0.12$ prolate deformation of $7^-$ and $8^-$ bands \cite{b2}. This effect shall also have direct impact on the prolately deformed $(\pi h_{11/2})^2$ alignment, which the shell-model framework can not intrinsically consider yet. Therefore, our calculated excitation energy of the $(\pi h_{11/2})^2$ alignment should be reduced by $\sim 0.6$ MeV as an amendment. By searching the $(\pi h_{11/2} \times \pi h_{11/2})^{I^{\pi}=10^+}$ pair in output wave-functions, the experimentally unobserved band beyond the $(\pi h_{11/2})^2$ alignment has been located in calculated spectrum, as demonstrated by red levels in Fig. \ref{spe}(b). The excitation energy of corresponding $10^+$ bandhead (i.e., the $(\pi h_{11/2})^2$ alignment) is $E\sim 5.2$ MeV. Thus, the $^{132}$Ba $(\pi h_{11/2})^2$-alignment energy is predicted to be around $4.6=5.2-0.6$ MeV, which agrees with the systematics of the alignment energy in light Ba isotopes \cite{neg-band-2,ps-10,131ba}.

\begin{table}
\caption{E3 transition rates (in W.u.) from the $I^{\pi}=10^+$ isomer with the $(\pi h_{11/2})^2$ alignment to states in $7^-$ and $8^-$ bands. The superscript ``$^{*}$" labels the main decay branch.}\label{e3}
\begin{tabular}{llcllccccccccccccccccccccccccccc}
					\hline\hline
$J^{\pi}_f$	&	B(E3)	&$~~~~~~$		&	$J^{\pi}_f$	&	B(E3)	\\
\hline													
$7^-$	&	0.008	&		&	$8^-$	&	0.145	\\
$9^-$	&	0.109$^{*}$	&		&	$10^-$	&	0.023	\\
$11^-$	&	0.001	&		&	$12^-$	&	$10^{-4}$	\\
					\hline\hline
				\end{tabular}
\end{table}

To probe electromagnetic properties of the $\sim 4.6$ MeV $(\pi h_{11/2})^2$-alignment state, we calculate its E2, M1, E3 decay to observed levels. Resultant E2 and M1 decay rates to yrast $8^+$, $10^+$ and $12^+$ states are all smaller than $10^{-13}$ W.u.. Such weak decays can be attributed to the parity conservation, which forbids E2 and M1 transition operators from scattering the negative-parity $h_{11/2}$ nucleon to other positive-parity orbits of 50-82 major shell. In other words, any state related to the $(\pi h_{11/2})^2$ alignment by strong E2 or M1 transitions must be constructed with two $h_{11/2}$ valence protons. However, yrast states, as well as most of low-lying positive-parity states in $^{132}$Ba, have few valence $h_{11/2}$ protons due to the large $\pi h_{11/2}$ single-particle energy. Thus, strong E2 and M1 decays from the $(\pi h_{11/2})^2$ alignment to these low-lying positive levels are absent, which may partially explain the inaccessibility of the $(\pi h_{11/2})^2$ alignment in observed level scheme of $^{132}$Ba. On the other hand, we obtain relatively strong E3 decays from the $I^{\pi}=10^+$ state with the $(\pi h_{11/2})^2$ alignment to states of $7^-$ and $8^-$ bands as shown in Table \ref{e3}. Such strong E3 decays are expected, because the $(\pi h_{11/2})^2$ alignment shares the prolate deformation with $7^-$ and $8^-$ bands. Based on above calculated decay rates, the $I^{\pi}=10^+$ level with the $(\pi h_{11/2})^2$ alignment is suggested to be a 0.511$\mu$s isomer with the main decay branch of $E_{\gamma}\sim 1.3$MeV to the $I^{\pi}=9^-$ level in the $7^-$ band.


\begin{figure}
\includegraphics[angle=0,width=0.48\textwidth]{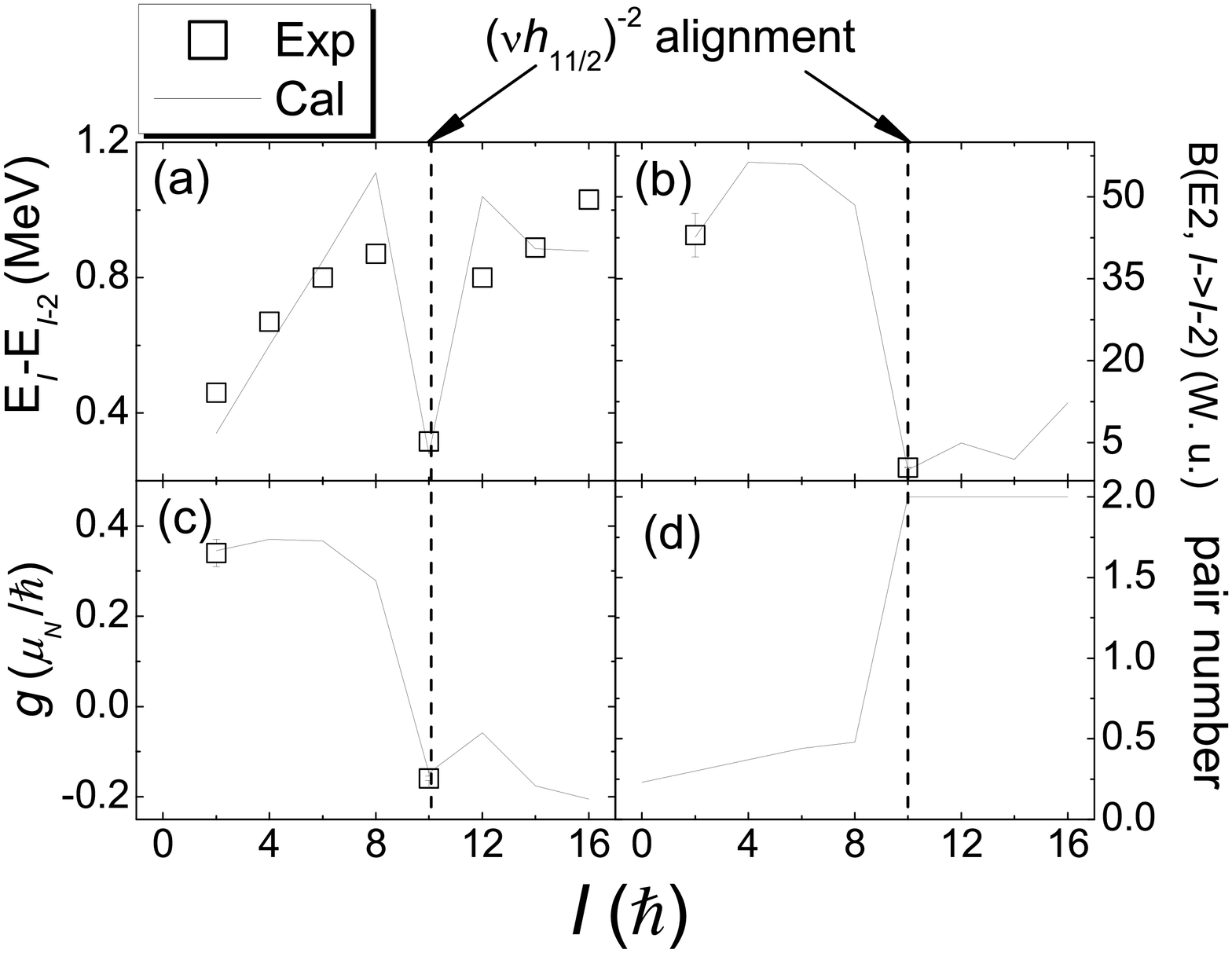}
\caption{$E_I-E_{I-2}$ (a), B(E2, $I\rightarrow I-2$) (b) and $g$ factors (c), as well as the sum of $5^-$ and $6^-$ pair numbers (d), of yrast states in $^{132}$Ba from experiments (Exp) \cite{ensdf} and our calculation (Cal). The yrast $(\nu h_{11/2})^2$ alignment is highlighted.}\label{yrast}
\end{figure}

Now, let's turn to the $(\nu h_{11/2})^{-2}$ alignment. The yrast $I^{\pi}=10^+$ isomer is the most typical $(h_{11/2})^{-2}$-alignment observation in $^{132}$Ba. We would like to revisit the yrast alignment before discussing $(\nu h_{11/2})^{-2}$ alignments in negative-parity bands. According to Fig. \ref{spe}, our calculation spectrally reproduces the excitation energy of the yrast alignment. To present this alignment more explicitly, we additionally compare calculated and experimental associated energies [$E_I-E_{I-2}$], reduced E2 transition rates [B(E2, $I\rightarrow I-2$)] and magnetic $g$ factors of the yrast band in Fig. \ref{yrast}. The backbending of $E_I-E_{I-2}$, the sharp dropping down of B(E2, $I\rightarrow I-2$) and $g$ factor at the $I^{\pi}=10^+$ isomer both experimentally and theoretically  evident the intruding of the $(\nu h_{11/2})^{-2}$ alignment in the yrast band. To observe this alignment at the wave-function level, we also calculate the expectation value of $5^-$ and $6^-$ pair numbers, and plot the sum of $5^-$ and $6^-$ pair numbers in Fig. \ref{yrast}(d). Correspondingly to the $(\nu h_{11/2})^{-2}$ alignment at $I=10\hbar$, two $5^-$ and/or $6^-$ pairs are suddenly excited as a representation of the $5^-\otimes 6^-$ coupling. In other words, the $(\nu h_{11/2})^{-2}$ alignment indeed is induced by the $5^-\otimes 6^-$ coupling. Therefore, we take the sudden increasing of $5^-$ and $6^-$ pair numbers by two as a signal of the $(\nu h_{11/2})^{-2}$ alignment.

\begin{figure}
\includegraphics[angle=0,width=0.48\textwidth]{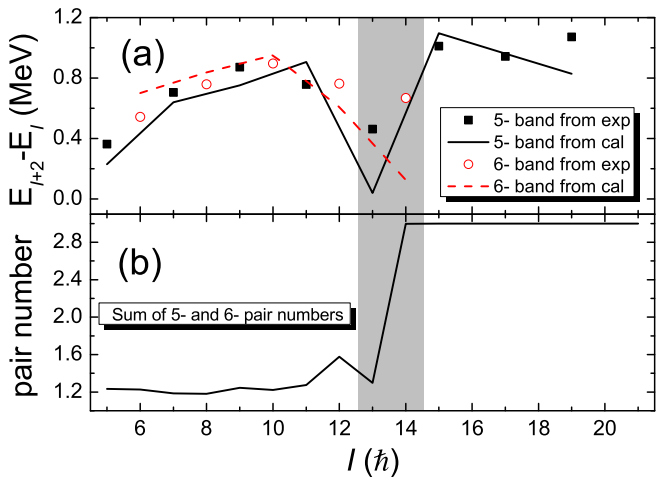}
\caption{(Color online) $E_{I+2}-E_I$ (a) and the sum of $5^-$ and $6^-$ pair numbers (b) in $5^-$ and $6^-$ bands. Abbreviations ``exp" and ``cal" correspond to the experimental data \cite{ensdf} and calculated results, respectively. The grey zone highlights the $(\nu h_{11/2})^{-2}$ alignment.}\label{56}
\end{figure}

The $(\nu h_{11/2})^{-2}$ alignment in $5^-$ and $6^-$ bands was proposed according to the band irregularity around $I=14\hbar$ in References \cite{neg-band-1,neg-band-2}. In Fig. \ref{56}(a), level spacings ($E_{I+2}-E_I$) also occur a sudden decrease around $I=14\hbar$, and more implicitly demonstrate this band irregularity in both experimental and calculated level schemes. We present the sum of $5^-$ and $6^-$ pair numbers in Fig. \ref{56}(b) correspondingly. Below $I=14\hbar$, only one $5^-$ or $6^-$ pair exists in $5^-$ and $6^-$ bands as expected by Ref. \cite{neg-band-1,neg-band-2}; beyond this point, two more $5^-$ and/or $6^-$ pairs, i.e., the $5^-\otimes 6^-$ coupling, are excited. The $(\nu h_{11/2})^{-2}$ alignment in $5^-$ and $6^-$ bands is confirmed. 

\begin{figure}
\includegraphics[angle=0,width=0.48\textwidth]{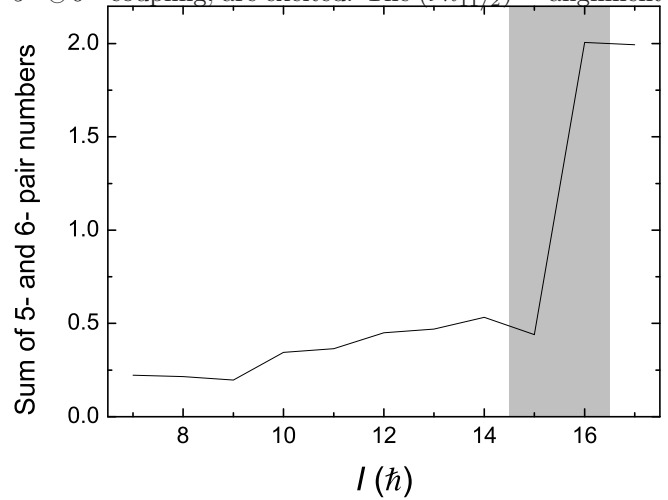}
\caption{Sum of $5^-$ and $6^-$ pair numbers in $7^-$ and $8^-$ bands. The grey zone highlight the potential $(\nu h_{11/2})^{-2}$ alignment.}\label{78}
\end{figure}

The $(\nu h_{11/2})^{-2}$ alignment is also expected in $7^-$ and $8^-$ bands around $I=16\hbar$ \cite{neg-band-2}, even though the spectral evidence was insufficient. We plot the sum of $5^-$ and $6^-$ pair numbers in $7^-$ and $8^-$ bands in Fig. \ref{78}, where the $5^-\otimes 6^-$ coupling is observed beyond $I=16\hbar$. Thus, we confirm the $(\nu h_{11/2})^{-2}$ alignment with oblate shape in $7^-$ and $8^-$ bands, although these two bands favor the prolate deformation.

To summarize, we adopt the pair-truncation of the Shell Model with negative-parity pairs to describe the $(h_{11/2})^2$ alignment in $^{132}$Ba. Our calculation is spectrally consistent with experiments. The $I^{\pi}=10^+$ state with the $(\pi h_{11/2})^2$ alignment is predicted to be an $E\sim4.6$MeV and $\tau\sim 0.5\mu$s isomer with relatively strong E3 transitions to $7^-$ and $8^-$ bands, which requires further experimental verifications. $(\nu h_{11/2})^{-2}$-alignment observations in both yrast band and negative-parity bands are also well reproduced by the $5^-\otimes 6^-$ coupling, which is suggested to be another representation of the $(\nu h_{11/2})^2$-alignment configuration. This oblately deformed alignment in prolate-favor $7^-$ and $8^-$ bands typically evidences the $\gamma$ unstability of $^{132}$Ba.

This work was supported by the National Natural Science Foundation of China under Grant No. 11305151. Discussion with Prof. Y. M. Zhao is greatly appreciated.

\end{document}